# One more experiment on "fast–light"

*Roberto Assumpção*[*]
*PUC-Minas, Av. Pe. Francis C. Cox, Poços de Caldas- MG  37701-355, Brasil*
*11 December 2003*

Contemporary observational and theoretical studies on the temporal nature of microscopic measurements renewed the discussion about the fundamental constants, leading to the possibility of light speed variation and superluminal pulse propagation. Gain assisted experiments using anomalous dispersion near an absorption line in atomic gas, a "fast – light" medium , seem to lead to a wave group velocity $v_G$ exceeding c, the vacuum speed of light; moreover, definition of the information velocity $v_i$ sets the question of interpretation of the three speeds: one view is that $v_i = v_G$ , but this violates Causality; another view is that $v_i = c$ in all situations, but this limits, *a priori*, the transport of information. Another view is that $v_i$, $v_G$ and c are distinct. This contribution follows the last possibility. A draft discussion on space-time is given.

The speed of light has so many facets [1] and is not exactly defined mainly due to two reasons: from one side, any pulse comprises a set of elementary waveforms, each one with a particular frequency; the second point is that laboratory tests mixes two beams – the one that is being measured and the one that is used to transport the information, that is, to establish the base-time.
Superluminal [2-5] detection can be based on anomalous dispersion [6], or the existence of a large discontinuity in the dispersion curve as it crosses an absorption band of a substance, so that longer wavelengths are more refracted than the shorter ones. If a light pulse of frequency $\omega$ and bandwidth $\Delta\omega$ enters a dispersive medium of refractive index $n(\omega)$, it's peak is supposed to propagate at the group velocity $v_G = c / n_G$ whereas each sinusoidal component travels at a phase velocity $v_{PH} = c / n(\omega)$; since the group index $n_G$ can be written in terms of the refractive index $n(\omega)$, $n_G \cong n(\omega) + \omega dn(\omega)/d\omega|_{\omega=\omega_o}$ , where $\omega_o$ is the central frequency of the wavepacket, in spectral regions where $n(\omega)$ decreases with $\omega$ ($dn(\omega)/d\omega < 0$) , the group index is less than the refractive index and can even become less than 1, resulting in a "fast–light" $v_G > c$ .
Current experimental arrangements [7] (Fig.1) use the technique of tuning a pulse near and far from the atomic resonance – $\omega_R$ – of a vapour. When $\omega \sim \omega_R$ , the pulse interacts with the "fast-light" medium and experiences an advancement; we refer to this velocity as $v_G$. Far from $\omega_R$ , there is no interaction and the velocity is $v_i$, usually taken as c, the speed of light in vacuum, under the argument that the "fast-light" medium is equivalent to vacuum when $\omega \neq \omega_R$. Alternatively, figure 2 pictures an experimental set-up based on tuning a "fast-light" optical medium on / off instead of tuning a laser near and far from the atomic resonance. This set-up discriminates the pulses propagating through vacuum ( information velocity $v_i$ ) from pulses propagating through the "fast-light" medium ( group velocity $v_G$ ); we hope that this can also help to distinguish a signal (or pulse) propagating through vacuum – $v_i$ from the speed of light in vacuum – c .

---

[*] Electronic address: assumpcao@pucpcaldas.br *;* assump@fem.unicamp.br

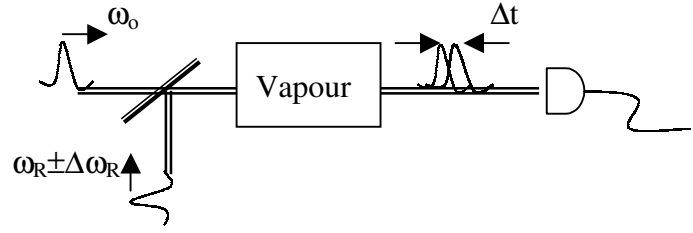

**Figure 1** Propagation of light pulses trough a dispersion medium: a coherently-prepared beam ($\omega_R$) tuned near ($\Delta\omega_R$) the atomic resonance is mixed with a beam ($\omega_o$) centred between the gain resonances, produces a pulse. $\Delta t$ is the "advancement" of this pulse relative to one propagating trough the cell when the lasers are tuned far from the atomic resonance, so that the path (trough vapour) is equivalent to vacuum (please see Ref.[7] for details).

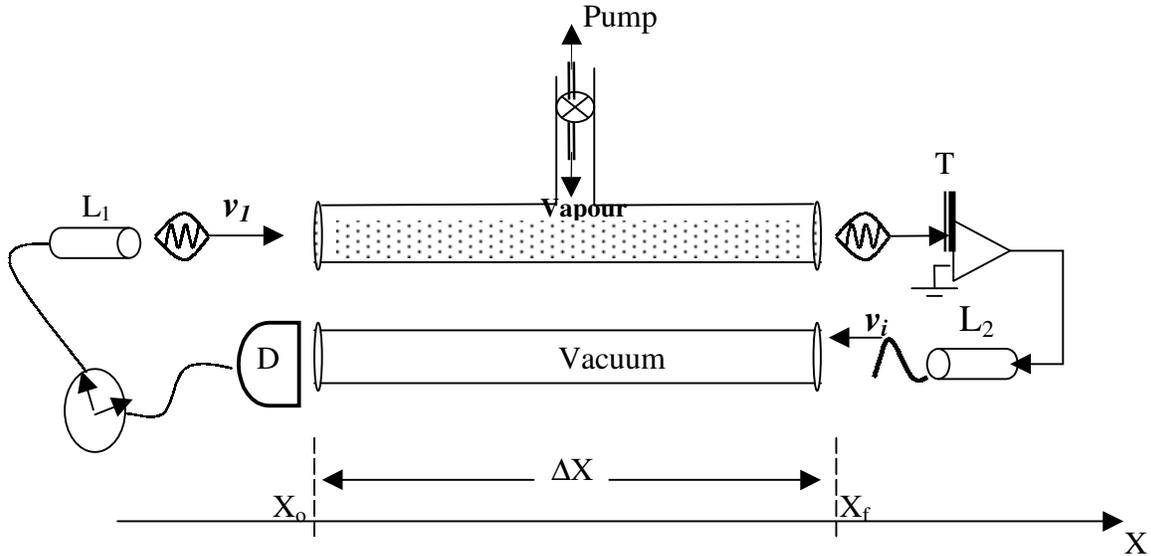

**Figure 2** An equivalent circuit : source $L_1$ starts detector D emitting a wave packet that propagates trough the upper arm at velocity $v_1$ . At T ($X_f$ ) the wave packet triggers the source $L_2$ that emits the signal $v_i$ trough the vacuum tube; this signal stops D ($X_o$). When the upper arm is filled with an atomic vapour, $v_1$ equals the group velocity $v_G$ and a superluminal regime is established. Pumping out the gas, $v_1$ equals the information velocity $v_i$ ; a balance between $v_G$ and $v_i$ is achieved by measuring the time interval registered by D .
For the superluminal regime we have:

$$\Delta t_{G,D} = \Delta t_G + \Delta t_i + \Delta t_{R,G} \quad (1)$$

where $\Delta t_{G,D}$ is the experimental time detected by "D", $\Delta t_G$ is the propagation time trough the vapour arm, $\Delta t_i$ is the propagation time trough the vacuum arm and $\Delta t_{R,G}$ is the time delay of the circuit; let this include the response time of "T" and "D".
For the vacuum regime we have:

$$\Delta t_{i,D} = \Delta t_i + \Delta t_i + \Delta t_{R,i} \quad (2)$$

where $\Delta t_{i,D}$ is the experimental time detected by "D", the first $\Delta t_i$ is the propagation time trough the upper arm (now evacuated), the second $\Delta t_i$ is the propagation time trough the vacuum arm and $\Delta t_{R,i}$ is the new time delay of the circuit, once the response time of "T" could be different for pulses $v_G$ and $v_i$; again, let this include the response time of "T" and "D". The measured difference $\Delta t_m$ between the detection times is:

$$\Delta t_m = \Delta t_G - \Delta t_i + (\Delta t_{R,G} - \Delta t_{R,i}) \quad (3)$$

Now we note that in this last expression, the experimental (measurable) quantities are $\Delta t_m$, $\Delta t_{R,G}$ and $\Delta t_{R,i}$; on the other hand, the superluminal state, if any, is due to the difference between $\Delta t_G$ and $\Delta t_i$. In other words, the last two terms assert the precision of the test, so that an exact result is just:

$$\Delta t_m \geq \Delta t_G - \Delta t_i \quad (4)$$

Reconsideration of equation (1) leads to:

$$\Delta t_{G,D} \approx \Delta t_G + \Delta t_i \quad (1')$$

This means that the experimental quantity $\Delta t_{G,D}$ is completely (exactly) determined in terms of the elapsed times $\Delta t_G$ and $\Delta t_i$; but $\Delta t_G$ and $\Delta t_i$ are not experimental quantities, they can't be – *strictu sensu* – directly measured. However, employing the space-time platform, that is, the definition of time in terms of velocity and length, the last equation can be written as:

$$\Delta t_{G,D} \approx \frac{\Delta x}{v_G} + \frac{\Delta x}{v_i} \quad (5)$$

where the terms on the right side represent a way to determine (not to measure) $\Delta t_{G,D}$. Now it comes the task of attributing significance to $\Delta t_{G,D}$; it turns out that this experimental quantity can be associated to the measurable value of the group velocity - $v_G$, which can be written as:

$$\Delta t_{G,D} \equiv \frac{\Delta x}{v_{G,D}} \quad (6)$$

where the subscript "D" has been added in order to emphasise that $v_{G,D}$ is the experimental value of the group velocity $v_G$, that is, the velocity that would be measured if, instead of the photodetector "D" calibrated in $s^{-1}$ we had employed a photoreceiver (namely, a space velocity detector) calibrated in m/s. By noting (Fig. 2) that the wavepacket $v_G$ flows in the vapour arm whereas information $v_i$ flows in the vacuum arm and since the length $\Delta x$ can be directly measured, Eq. (6) into (5) gives:

$$\frac{1}{v_{G,D}} \approx \frac{1}{v_G} + \frac{1}{v_i} \quad (7)$$

implying that the group velocity can only be determined in terms of the information velocity:

$$v_G \approx \frac{v_{G,D}}{1 - \frac{v_{G,D}}{v_i}} \quad (8)$$

Now it is possible to examine the role of c, $v_i$ and $v_G$:
One view is to insert $v_i = v_G$ into Eq. (8); this gives a superluminal interpretation ($v_G \sim 2 v_{G,D}$) once the group velocity appears as twice its experimental value, somehow violating causality. But this is equivalent to an experimental set-up (Fig.2) in which the vacuum arm is filled with atomic vapour – the "fast-light" medium – so that the wavepacket $v_G$ carriers its own information, giving rise to a subluminal detection or sensation ($v_{G,D} = \frac{1}{2}v_G$). Since this does not correspond to the actual experiment, both the interpretation ($v_i = v_G$) and the sensation ($v_{G,D} = \frac{1}{2} v_G$) are misleading.

The second view is to insert $v_i = c$ into equation (8); however, equations (7) and (8) are written in the velocity-space ( Boltzmannian sense) platform whereas c belongs to the space-time ( Newtonian and Einstenian sense) platform. In fact, c appears as a space-time synthesiser, so the dimensionless ratio $v_{G,D} / v_i$ is incompatible with a $v_{G,D} / c$ interpretation by construction of the argument leading to the existence of the information velocity $v_i$. Indeed, the numerical value bellow $v_{G,D}$ in Eq. (8) is not arbitrary, once it defines a basis (platform) in terms of which $v_G$ is being measured. In order to transform from one platform to the other, it is necessary to measure $v_i$ that, as $v_G$, is unknown in Eq. (8).

For the vacuum regime, Eq. (2), we have:

$$\Delta t_{i,D} = \Delta t_i + \Delta t_i + \Delta t_{R,i}$$

where the first $\Delta t_i$ represents the flow trough the "vapour arm" ( now evacuated) and the second $\Delta t_i$ the flow trough the vacuum arm (Fig.2) ; again, $\Delta t_{i,D}$ and $\Delta t_{R,i}$ are the experimental (measurable) quantities; since the last asserts only the precision of the test, an exact result is:

$$\Delta t_{i,D} \approx \Delta t_i + \Delta t_i \quad (9)$$

Following the arguments above, we can write:

$$\Delta t_{i,D} \approx \frac{\Delta x}{v_i} + \frac{\Delta x}{v_i} \quad (10)$$

where the first term in the right side of equation (10) is the "group velocity" when the laser is tuned far from resonance, while the second term corresponds to the flow ( of information) trough the vacuum arm; therefore, $\Delta t_{i,D}$ represents the measured value of the first term:

$$\Delta t_{i,D} \equiv \frac{\Delta x}{v_{i,D}} \quad (11)$$

again, the subscript "D" has been added in order to emphasise that $v_{i,D}$ is the experimental value of $v_i$. Thus we have:

$$\frac{1}{v_{i,D}} \approx \frac{2}{v_i} \quad (12)$$

Perhaps the most striking feature of ordering objects is the outcome with which they can be characterised; for instance, one consequence of the arrangement of the title is the association between the words light and fast. There seems that light is fast but these objects (words) are just relative so those words (objects) should be balanced among themselves or between similar entities. In fact, Eq. (8) balances two velocities and says that the faster the information velocity is, more and more the experimental value approaches the true one, or better the result obtained for the group velocity. This is quite a reasonable result.

On the other hand, the vacuum regime represents an experimental set up where information is balanced by itself and Eq. (12) says that if $v_i = c$, $v_{i,D} \sim c/2$. But this contradicts the experimental fact that the speed of light in vacuum is c. Thus we are led to admit that it is $v_{i,D}$, the experimental value of $v_i$, that equals the speed of light in vacuum, so that $v_i \sim 2c$. Inserting this rough, though reasonable result into Eq. (8), we have:

$$v_G \approx \frac{v_{G,D}}{1 - \frac{v_{G,D}}{2c}} \quad (8')$$

Note that this equation is written in the velocity-space platform; the quantities $v_{G,D}$ and 2c are known from experiment so that, the fundamental problem of "velocimetry", reduces to the problem of establishing a velocity scale. That is, the experimental quantities that are (independently) measured in determining an unknown velocity ($v_G$) are quantities of velocities rather than lengths and times. In technical terms (Fig.2), this corresponds to a velocity measurement employing a space velocity detector calibrated in m/s and located at the end ($X_f$) of the upper arm (note that the lower arm is unnecessary in this case) instead of a photo-detector "D" calibrated in $s^{-1}$. In theoretical terms, this corresponds to a transformation [8] from space-time to a velocity-base given by:

$$\frac{1}{v_{G,D}} \rightarrow \frac{1}{v_G} + \frac{1}{v_i} \quad (7')$$

where the left term can be determined in space-time,

$$\frac{1}{v_{G,D}} = \frac{\Delta t_{G,D}}{\Delta x} \quad (6')$$

whereas the right side terms in Eq. (7') are written in the velocity-space. Though measurements consist of a set of dimensionless ratios, it seems that no one has ever measured a velocity in the manner described above. The c velocity scale is established by correcting the readings of a "velocimeter" to what they would be if the motion were ideal, that is, static $v_{G,D} / v_i \sim 0$. In short: a superluminal effect can be observed, depending on how fast information is transported and is a result related to the fact that velocities are underestimated in space-time. This effect is due to the limited value of $v_i$. In order to see that it is $v_i$ and not c the responsible for the spatiotemporal structure – the causal connection – let us first consider Fig. 2 and then equation (8).

From the experimental point of view, a measurement is completed when information is inside a detector; according to Fig.2, detector "D" is fixed at Xo and experiences two moments: the moment the wavepacket $v_I$ is launched (ON) and the moment the information $v_i$ arrives (OFF); these two moments give rise to a single time $\Delta t_m$ (Eq. (3)). One moment defines a point in time, but not a point in space-time. In German this sounds more robust: moment = zeitpunkt. This suggests that point Xo belongs to space-time but point $X_f$ belongs to the velocity-space, so that Xo is a point of Simultaneity whereas $X_f$ is a point of Synchronicity. Synchronism, or the causal connection between both points is achieved, experimentally, by $v_i$ and not by c. c is just the result (or the synthesis) of the connection between space and time; the responsible is $v_i$.

From the conceptual point of view, Eq. (8) and more precisely Eq. (7) say that the measured value of the group velocity is the medium value obtained by the harmonic relation between the actual flow velocity and the information velocity. The "collapse" of the two velocities, $v_G$ and $v_i$, leading to the measured value $v_{G,D}$ embodies the measurement uncertainty. A plot of equation (8) in units of c is given in figure 3.

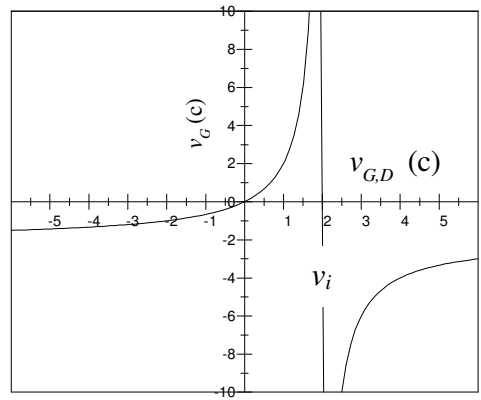

**Figure 3** Group velocity in terms of the measured value; the line crossing 2c is $v_i$.

Adopting positive velocities as those signals starting from the source point Xo in Fig.2, we note that the vertical axis in Fig.3 is distorted due to the finite value of $v_i$ and "delayed" according to Eq. (1'); in fact, the velocities in Eq. (7) should be associated to the times in equation (1'). Formally, the reciprocal velocity vector addition $1/v_G + 1/v_i$ is completely defined by the coordinates of its end point, $1/v_{G,D}$. We can therefore speak of this end point, instead of the reciprocal vectors themselves, and describe the measurement problem as that of determining the way these representative end points are distributed in the velocity space. On the opposite, in the preceding paragraphs we noted that $v_{G,D}$ is completely determined by its space-time coordinates – Eq. (6' ); therefore, in terms of temporalities, this seems to imply that the arrowhead of the reciprocal vector $1/v_{G,D}$ is centred on Past, or that this vector points to past, so that the useful portion of Fig.3, the first quadrant ($v_{G,D} < v_i$ ), can be associated to Past. The limitation of our observation is thus translated as a delay or in the fact that all measurement is a detection of what has passed. But the measurement problem can be stated in another way: employing Eq. (8) and its counterpart, Eq. (1' ), together with the definition of velocities in space-time,

$$v_K \equiv \frac{\Delta x}{\Delta t_K} \quad (13)$$

where K labels the distinct velocities and times,

and taking $\frac{\Delta x}{\Delta t_{i,D}} = c$ , the experimental value of the speed of light in vacuum,

it is straightforward to show that:

$$\Delta x^2 - 4c^2 \Delta t^2 = 0 \quad (14)$$

which could be generalised to,

$$(x - x_o)^2 + (y - y_o)^2 + (z - z_o)^2 - 4c^2(t - t_o)^2 = 0 \quad (15)$$

where $\Delta t = t - t_o$ is formally associated to $\Delta t_i$ , so that one could refer to the four dimensional point (x, y, z, t) as an "event" in space-time; note, however, that an event is defined (a measurement is completed) only when/where information is inside a detector located in the space-time point ($x_o$, $y_o$, $z_o$, $t_o$).

According to special theory of relativity, no signal can cause an effect outside the light cone [9], defined as the space-time surface on which light rays emanate from the source; here we note that according to Fig.3, equation (15) is the "usual" space-time counterpart of the velocity-space equation (8), provided the role of the speed of light is played by the information velocity in the last term, that is, the temporal line element $4c^2 dt = v_i^2 dt$. Further, since causality is intrinsic, by construction ( Fig.2; equation (8)), we may ask that equation (15) represents a causal cone that will coincide with the light cone only under special circumstances. Let us examine this cone in terms of the remaining temporalities, Present and Future, that necessarily should appear in Fig.3 .

In order to reduce the delay, that is, to approach the measured value to the actual one, one could assume, as usual, that $v_G$ equals $v_{G,D}$. But this means that uncertainty relations ( such as equation (4) ) do not hold. A more sounded physical assumption is to increase $v_i$ or equivalently reduce $v_G$. This is equivalent to a translation of the fourth quadrant curve to the $v_G = 0$ point, generating a "cone" centred on this point. Since this corresponds to a perfect measurement (no delay), the fourth quadrant curve should be associated to Present. In other words, the light cone collapses Present to a point in space-time whereas the causal cone reveals the full curve (a surface in velocity or momentum-space ) beyond the information velocity. This limit ($v_G = 0$) appears as a classical (Newtonian) particularity, where the actual value of the velocity corresponds to the measurement, an "effect" due to the "infinite" value of $v_i$, that is, an ideal (static) measurement. Causality is preserved for all positive velocities in this plot, provided $v_i > v_{G,D}$. On the other hand, a faster than $v_i$ signal appears either as positive velocity ($v_{G,D}$) in the fourth quadrant or as a negative one in the third. The former lacks physical importance mainly due to the reason that it implies an effect preceding its cause or an exchange between Past and Future. Note that this is equivalent to measure a light pulse employing, for instance, a sound wave as $v_i$. The latter, however, has a physical meaning if we adopt negative velocities as those signals arriving in the source point $X_o$ in Fig.2. There are two such signals: those coming from point $X_f$, that is, the information $v_i$ and those yet unemitted signals, that is, signals that will come into the apparatus in near future. Indeed, since it is the value of $v_i$ that determines the measurable limit at 2c and those detectable signals are restricted to $v_i \geq |v_G|$, it can be argued that point $X_f$ in Fig.2 belongs to (far) future, so that information is always coming from future. It follows that the experimental process pictured in Fig.2 discriminates temporalities and hence the familiar notion of "absolute" time, revealing a temporal dependence on the relative magnitude and sign of the signals. This is accounted for by $v_i$, the responsible for the space-time causal connection and not by c, the synthetic result of causality.

To conclude, we note that: 1. The possibility that the fundamental space-time constant c may vary exists but, detection of such a variation is limited in space-time (Michelson-Morley type experiments) due to the finite value of the information velocity; 2.Superluminal experiments do not test causality; they verify the uncertainty principle; and it holds; 3. Due to the uncertainty principle and to the constant experimental value c, light, itself is superluminal. 4. If one were to prove mentally that $c < v_i \sim 2c$, then the standard derivation of special relativity would have to be revised; 5. The distinction among $v_G$, $v_i$ and c do not contradict variable speed of light theories, superluminal phenomena, neither shakes the overall validity of special relativity; it could, however, nullify derivations of the Lorentz transformation based on the invariance of the photon velocity.

#

[1] Ellis, G F R and Uzan. J P ; gr-qc/0305099
[2] Garret, C G B and McCumber, D E ; Phys. Rev. A 1, 305-313 (1970)
[3] Casperon, L and Yariv, A ; Phys. Rev. Lett. 26, 293-295 (1971)
[4] Chiao, R Y ; Phys. Rev. A 48, R34 – R37 (1993)
[5] Wang, L J , Kuzmich, A and Dogariu, A ; Nature, 406, 277-279 (2000)
[6] Jenkins, F A and White, H E, Fundamentals of Optics, 4[th] Ed., sec. 23.4 (McGraw-Hill, São Paulo, 1981)
[7] Stenner, M D, Gauthier, D J and Neifeld, M A ; Nature, 425, 695-698 (2003)
[8] Assumpção, R ; physics/0305037; Alternative Theories, in Tenth Marcel Grossmann Meeting, July, 2003.
[9] Symon, K R ; Mechanics, 3[rd] Ed., sec. 13.3-13.5 (Addison-Wesley, London,1972)

**Acknowledgements** I'm grateful to Daniel Gauthier for bringing his works at the Duke Physics site to my attention; a far past discussion with Waldir A. Rodrigues Jr., who also contributed to the site, is also acknowledged.